\documentclass[conference, 11 pt]{IEEEtran}
\IEEEoverridecommandlockouts
\usepackage{cite}
\usepackage{amsmath,amssymb,amsfonts}
\usepackage{algorithmic}
\usepackage{graphicx}
\usepackage[singlespacing]{setspace} 
\usepackage{textcomp}
\usepackage{xcolor}
\usepackage{adjustbox}

\usepackage{booktabs} 

\usepackage{multirow}
\def\BibTeX{{\rm B\kern-.05em{\sc i\kern-.025em b}\kern-.08em
    T\kern-.1667em\lower.7ex\hbox{E}\kern-.125emX}}

\begin{document}

\title{Image registration based automated lesion correspondence pipeline for longitudinal CT data\\
}

\author{\IEEEauthorblockN{Subrata Mukherjee\IEEEauthorrefmark{1}, 
Thibaud Coroller\IEEEauthorrefmark{2}, 
Craig Wang\IEEEauthorrefmark{3}, 
Ravi K. Samala\IEEEauthorrefmark{1}, 
Tingting Hu\IEEEauthorrefmark{1}, 
Didem Gokcay\IEEEauthorrefmark{1}, \\ 
Nicholas Petrick\IEEEauthorrefmark{1}, 
Berkman Sahiner\IEEEauthorrefmark{1}, 
Qian Cao\IEEEauthorrefmark{1}}
\IEEEauthorblockA{\IEEEauthorrefmark{1}U.S. Food and Drug Administration, \\Center for Devices and Radiological Health, Office of Science and Engineering Labs, \\Division of Imaging, Diagnostics, and Software Reliability, Silver Spring, MD}
\IEEEauthorblockA{\IEEEauthorrefmark{2}Novartis Pharmaceutical Corporation, New Jersey, USA}
\IEEEauthorblockA{\IEEEauthorrefmark{3}Novartis AG, Switzerland}
}

\maketitle

\begin{abstract}
Patients diagnosed with metastatic breast cancer (mBC) typically undergo several radiographic assessments during their treatment. Accurately detecting and monitoring individual lesions over time, is crucial for making informed clinical decisions. mBC often involves multiple metastatic lesions in different organs, therefore it is imperative to accurately track and assess these lesions to gain a comprehensive understanding of the disease's response to treatment. Computerized analysis methods that rely on lesion-level tracking have often used manual matching of corresponding lesions, a time-consuming process that is prone to errors. This paper introduces an automated lesion correspondence algorithm designed to precisely track both targets' lesions (mBC lesions that are actively monitored and tracked) and non-targets' lesions (abnormalities that are not currently of primary focus) in longitudinal data. As part of a collaboration between U. S. Food and Drug Administration and Novartis Pharmaceuticals, we demonstrate the applicability of our algorithm on the anonymized data from two Phase III trials, MONALEESA-3 and MONALEESA-7. The dataset contains imaging data of patients for different follow-up timepoints and the radiologist annotations for the patients enrolled in the trials. Target and non-target lesions are annotated by either one or two groups of radiologists. To facilitate accurate tracking, we have developed a registration-assisted lesion correspondence algorithm. The algorithm employs a sequential two-step pipeline: (a) Firstly, an adaptive Hungarian algorithm is used to establish correspondence among lesions within a single volumetric image series which have been annotated by multiple radiologists at a specific timepoint. (b) Secondly, after establishing correspondence and assigning unique names to the lesions, three-dimensional (3D) rigid registration is applied to various image series at the same timepoint.  Registration is followed by ongoing lesion correspondence based on the adaptive Hungarian algorithm and updating lesion names for accurate tracking.  This iterative algorithm is then applied to all timestamps and extended across multiple timepoints for a given patient to ensure precise temporal tracking of targets and non-targets. Validation of our automated lesion correspondence algorithm is performed through triaxial plots based on axial, sagittal, and coronal views, confirming its efficacy in matching lesions.
\end{abstract}


\section{Introduction}
This paper presents work from a collaboration between the U. S. Food and Drug Administration and Novartis, where we retrospectively utilize anonymized data from Phase III clinical trials, MONALEESA-2, 3 and 7, to investigate novel prognostic and predictive factors for metastatic breast cancer patients \cite{coroller2023methodology,coroller2023multi}.  The clinical trials were originally conducted to evaluate the efficacy and safety of ribociclib as an investigational drug in combination with endocrine therapy leading to the approval of ribociclib as a targeted therapy for metastatic breast cancer (mBC), specifically mBC characterized by hormone receptor-positive (HR+) and human epidermal growth factor receptor 2-negative (HER2-) status \cite{tripathy2018ribociclib,slamon2018phase,hortobagyi2016ribociclib}. Metastases are a leading cause of cancer-related mortality \cite{chaffer2011perspective}, and requires longitudinal imaging to track metastases for effective patient management. The imaging subcomponent of our larger collaboration aims to employ CT scan images at various time points to evaluate tumor progression and guide treatment decisions. The CT images in our dataset were read  by either one or two radiologists where they annotated metastatic lesions as part of a blinded independent central review process. 

Computerized analysis of this longitudinal CT data, annotated by different radiologists, presents challenges, including the appearance of new lesions, the disappearance of responding lesions, differnt annotations by different radiologists and across timepoints, and variations in the field of view (FOV) due to different scan series. Manual sorting of lesions and correspondences acoss longitudinal CTs is highly labor and time intensive. Additionally, metastatic cancers may exhibit diverse responses across multiple lesions \cite{humbert2020dissociated,wang2012study,huff2023performance}. Imaging can vary over time due to different patient alignment with respect to the scanner leading to different image scan series and thereby different field of views. The anatomy of patients may also change between time points. Moreover, annotations by different radiologists often leads to discrepancies in lesions and lesion naming, making manual matching and tracking of lesions challenging. Therefore, we developed an automated lesion correspondence algorithm capable of effectively tracking distinct lesions throughout the longitudinal imaging study, while addressing the aforementioned complexities and challenges.

Previous literature \cite{an2021robust,xu2011automated,santoro2021development} have addressed lesion matching through registration. However, registration approaches alone do not  fully capture the complexities of our mBC datasets. Our contribution in this work lies in addressing the following key challenges: establishing lesion correspondence despite annotation discrepancies among radiologists, handling multiple timepoints as well as different image scan series within each time point, and tracking non-target lesions in addition to the RECIST \cite{muenzel2012intra} target lesions.
Our proposed algorithm utilizes the Hungarian (also named Kuhn-Munkres) algorithm \cite{munkres1957algorithms,mills2007dynamic}, along with a distance threshold and 3-dimensional (3D) rigid registration \cite{marstal2016simpleelastix}, in a sequential manner. The algorithm dynamically updates the correspondence assignment process, ensuring consistent naming for matched lesions and new names for unmatched ones. We employ a combinatorial optimization technique to address the assignment challenge involving lesion centroid coordinates from distinct radiologists working on the same scan series at a specific timepoint. This task is carried out independently for both targets and non-targets, considering each lesion as a vertex and their respective pairings as edges (details in the section~\ref{algorithm}). This graph-based correspondence approach is systematically applied to scans evaluated by two different groups of radiologists. Subsequently, we perform volumetric registration across different scan series originating from the same timepoint. Following this, the correspondence algorithm is reapplied, leading to updates in the nomenclature of targets and non-targets. This iterative process is carried out for each timepoint and extended to multiple timepoints. Depending on the number of image series and timepoints, volumetric registration may need to be run multiple times. We employed rigid registration, which offers faster results compared to other deformable volumetric registration techniques \cite{larrey2006optimal}.

The rest of the paper is organized as follows: In section~\ref{datagen}, we describe our study population and the variabilities in the MONALESSA dataset. In Section~\ref{algorithm}, we outline our correspondence and registration algorithms. Section~\ref{results} shows the results of the algorithm in the form of triaxial plots on a sample of patients. Finally, Section~\ref{conclusion} concludes the preliminary testing,  and discusses potential future work.

\section{CT Imaging Data}\label{datagen}

The effective sample from MONALEESA-3 and MONALEESA-7 encompass approximately 300 patients. Tumor evaluations occur initially at baseline (screening), followed by every 8 weeks during the first 18 months, and then every 12 weeks thereafter, until disease progression,  patient passing, or withdrawal from the study, whichever comes first. We specifically focused on the first two timepoints, namely Screening and Week 8 in this study. Target lesions were annotated by either a radiologist from one radiologist group, or two independent radiologists from two groups, adhering to the Response Evaluation Criteria in Solid Tumors (RECIST) criteria v1.1 \cite{muenzel2012intra}. The same radiologists also provided two-dimensional (2D) segmentations of the RECIST target lesions. RECIST non-target lesions were annotated, but not segmented. Each volumetric scan consists of a three dimensional (3D) image with dimensions of approximately $512 \times 512 \times \sim 100$ voxels. 

\begin{figure*}[!h]
\centering
  \includegraphics[width=1\textwidth,height=3in]{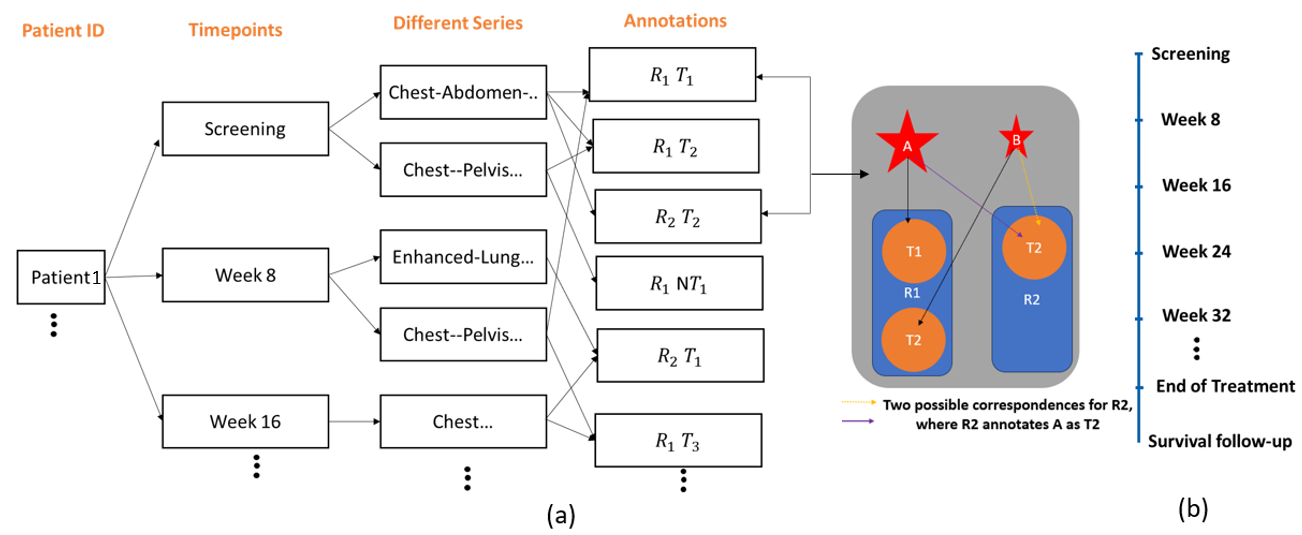}
  \caption{(a) Overview of the complex architecture in our MONALEESA datasets reflecting the discrepancy in annotation across radiologists, (b) a possible scenario for data availability throughout the patient's participation in the clinical trial. Here 'T' represents targets and 'NT' represents non-targets. }\label{fig1}
\end{figure*}

Figure~\ref{fig1}(a) shows the complex structure of our MONALEESA dataset where an arbitrary patient is typically inspected at multiple timepoints. At each timepoint, there can be one or multiple CT scan series with each CT series annotated by either one or two radiologists families ($R_{1}$ and $R_2$ are two different families annotating lesions as shown in figure ~\ref{fig1}(a)). Figure ~\ref{fig1}(b) reflects a possible scenario of data availability during a typical patient journey in the trial. 

\begin{figure}[!h]
\centering
	\includegraphics[width= 3 in,height= 3 in]{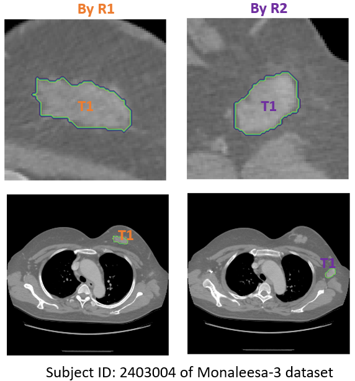}
	\caption{Example with two radiologist families marking the same lesion. Left column - in top zoomed axial and bottom axial orientation, radiologist (R1) marked one target lesion as T1. right column- in top zoomed axial and bottom axial orientation, radiologist (R2)  annaoates a different target as T1.}\label{fig2}
\end{figure}

A problem that arises in the analysis of the dataset is that there is no reliable lesion correspondence among the different radiologists. Figure ~\ref{fig1}(a) illustrates such discrepancies where one radiologist family (R1) may designate one lesion (A) as target 1 (T1) whereas another radiologist family (R2) may annotate the same lesion as target 2 (T2).  This leads to a swapping of lesion correspondence among radiologists (as shown in right hand side figure~\ref{fig1}(a)). Furthermore, as we progress to later weeks, new lesions may appear, or existing lesions may disappear.
Consequently, there is a clear need to establish unique lesion correspondences of lesions with and across radiologist and withing and across timepoints. Figure~\ref{fig2} illustrates a situation in which one radiologist (R1) designates one lesion as 'T1' during the screening, while another radiologist annotates a different lesion as 'T1' for the same timepoint.

\begin{figure}[!htbp]
\centering
	\includegraphics[width= 3 in,height= 2.3 in]{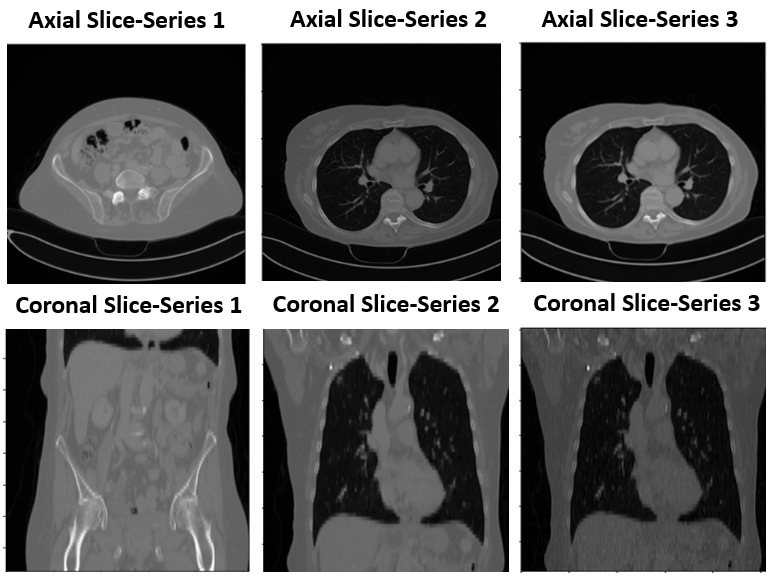}
	\caption{Example with different anatomical sections at the same timepoint. (a) Axial orientation (in top) and coronal orientation (in bottom) of three different scan series at the Week 8 timepoint or a subject.  Thsi variaotion in the series shows the need for registration because of the different anatomy scanned.}\label{fig3}
\end{figure}

Registration is needed as imaging across different positions of a patient leads to different scan series, each capturing a different range and section of the anatomy. Figure~\ref{fig3} highlights the necessity for performing registration, as there can be different image scan series with varying anatomical sections at a given follow-up timepoint. Figure~\ref{fig3} shows three different scan series with markedly different sections of a patient's anatomy at Week 8. 

%\clearpage
%\vspace*{0.20in}
%\hspace{1cm}\\[5ex]

These scenarios underscore the importance of developing a registration-assisted lesion correspondence algorithm for accurate tracking and unique naming of both targets and non-targets.

\section{Automated lesion matching algorithm}\label{algorithm}
We describe the Registration based Automated Lesion Matching and Correspondence (RALMAC) algorithm here. Initially, the lesion matching algorithm is applied to establish correspondence between lesion centroids, for both target and non-target lesions, from two separate radiologists for the same image series at a specific timepoint. Subsequently, the lesion matching algorithm is employed to determine correspondence across different series within a specific timepoint and across various timepoints.

\begin{figure}  
\centering
	\includegraphics[width= 3.5 in]{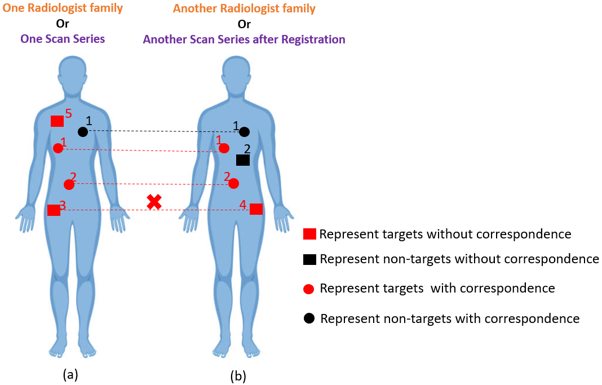}
	\caption{Schematic showing unique lesion matching concept. Target lesions are shown in red and non-target lesions in black. The matched lesions are inidcated by circles whereas unmatched lesions are indicated by squares.}\label{fig4}
\end{figure}

\textbf{Automated Lesion Matching and Correspondence Algorithm.} This algorithm is rooted in the Kuhn-Munkres algorithm \cite{munkres1957algorithms,mills2007dynamic} popularly known as the Hungarian algorithm. It employs combinatorial optimization to efficiently match lesion centroids based on their spatial coordinates. The objective is to find one-to-one correspondence between two sets of lesions by minimizing a cost function based on the Euclidean distance between lesion centroids in the two sets. For a particular image series, the two sets correspond to data frames containing lesion annotations from two different radiologists. In the case of different image series, the sets represent data frames from different scan series after registration.  The algorithm's working principle is illustrated in Figure~\ref{fig4}. and finds correspondence separately among target and non-target lesions. The centroids of the target lesions are marked in red, while those of non-target lesions are marked in black (in figure~\ref{fig4}). The Hungarian algorithm operates on the distances between all lesion centroids in these groups. 

The algorithm treats the correspondence problem as a bipartite graph, denoted as $G = \{R_1, R_2, E\}$ as illustrated in figure~\ref{fig4}. Each partition of the graph (a and b in figure~\ref{fig4}) $R_1 = \{r_{1,1}, \ldots, r_{1,i}\}$ represents target centroids (in red in figure~\ref{fig4}) of one radiologist / one image series  whereas $R_2 = \{r_{2,1}, \ldots, r_{2,j}\}$ represents target centroids (in red in figure~\ref{fig4}) of another radiologist / another image series. The same holds for the non-target groups (marked in black in figure~\ref{fig4}). The edge E indicates a match between the two sets (as marked by dotted line between matches). Unmatched lesions (lesion 5 and 2 in figure~\ref{fig4}) are assigned new names.  To enhance matching accuracy, an additional distance threshold is applied. Lesion pairs with distances exceeding this threshold are considered unmatched, even if the Hungarian algorithm suggests a correspondence based on the cost function. The distance threshold is 40 mm for the target lesions and $50$ mm for the non-target lesions.  This scenario is observed between lesion 3 and 4 in figure~\ref{fig4}  as the Hungarian alorithm suggest the lesions are matched but our distance threshold defines them as unmatched and assigns new names. After establishing correspondence, the algorithm assigns names to target lesions as $\{G_1, G_2, \ldots, G_j\}$ and to non-target lesions as $\{NG_1, NG_2, \ldots, NG_i\}$ with j and i unique target and non-target lesions after matching.

\begin{figure*}[!t]  
\centering
	\includegraphics[width=0.9\textwidth,height=3.5in]{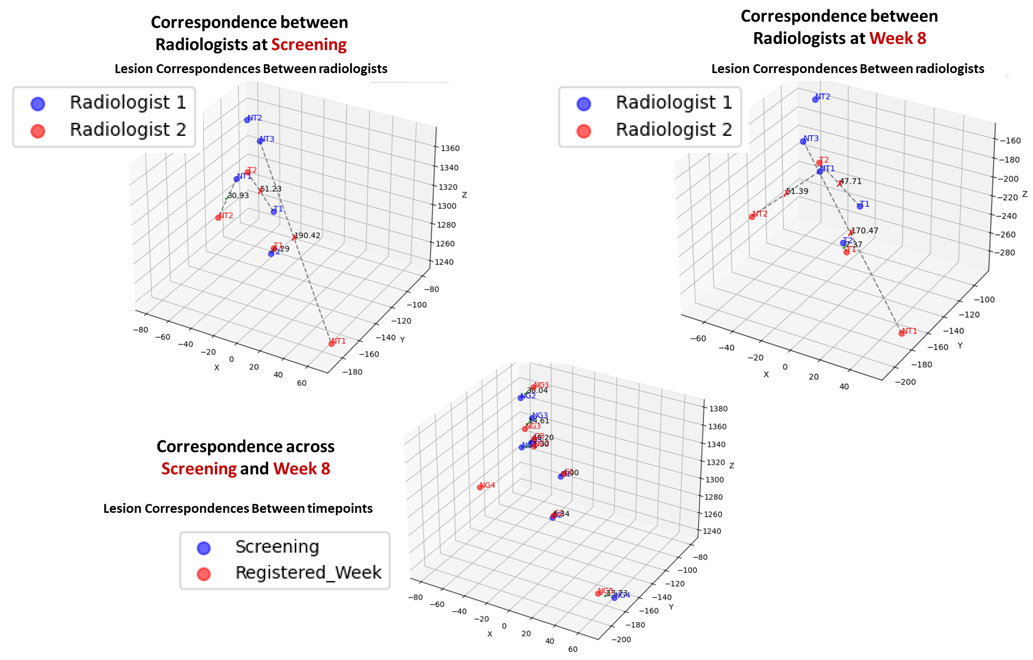}
	\caption{Top left upper plot shows the application of correspondence algorithm to match lesions across radiologists at screening. Top right plot shows the application of correspondence between radiologists at Week 8. Bottom plot shows the correspondence across the timepoints post registration.Lesion marked by radiologist 1 are in blue and those marked by radiologist 2 are in red (for the top two plots). In the bottom plot, lesions at 'Screening' time point are in blue and those at time point 'Week 8' are in red.}\label{fig5}
\end{figure*}

For an example, the algorithm first finds correspondence among radiologist families in the scan series at Screening ( top left plot of figure~\ref{fig5}). Then,repeats this process for the Week 8 scan series (top right plot). Finally, after volumetric registration across timepoints (Screening and Week 8), the algorithm establishes correspondence between lesions from Screening and Week 8 (shown in bottom plot of figure~\ref{fig5}) .

\textbf{Volumetric Rigid Registration.}
Volumetric (3D) rigid registration is a pivotal process in medical imaging analysis, facilitating the alignment of two 3D image datasets to enable spatial correspondence. This technique is instrumental in a multitude of clinical applications, such as monitoring disease progression, formulating treatment strategies, and conducting comparative analyses of anatomical structures across different time points or between patients. The Simple Image Toolkit (SITK) \cite{marstal2016simpleelastix} provides a robust framework for implementing 3D rigid registration, which we applied here to align different image series from a specific time stamp or between varying stages such as screening and a subsequent 8-week follow-up. 

\begin{figure}  
\centering
	\includegraphics[width= 3.5 in]{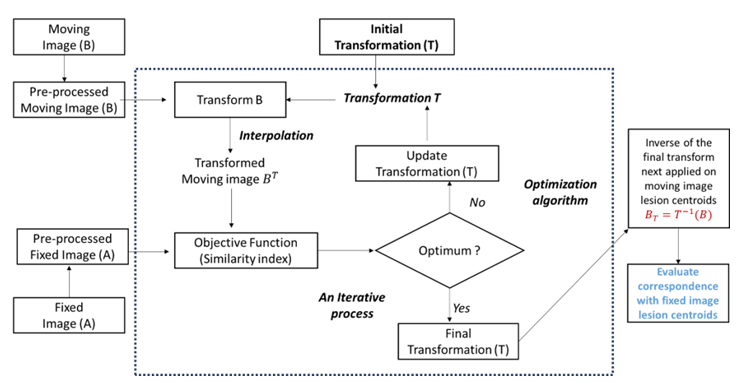}
	\caption{Schematic of the volumetric rigid registration framework based on the transformation approach adopted in the paper.}\label{fig6}
\end{figure}

Steps of the SITK based rigid registration is depicted in figure~\ref{fig6}. In rigid registration, the image series designated as the 'fixed' image serves as the benchmark, while the 'moving' image series is transformed to align with the fixed one. When performing registration across different timepoints, typically, the image at the screening phase is kept fixed, and subsequent images, like those from Week 8, are adjusted accordingly. 

Prior to registration, both fixed and moving images undergo preprocessing, including binary thresholding and connected component analysis. This step isolates the pertinent anatomical features while eliminating extraneous elements such as air pockets in the images. The rigid transformation, denoted by $T(S)$ maps position \( \mathbf{S} = [x, y, z]^T \) in fixed image (A) to corresponding position in moving image (B). Transformation T(X) has six degrees of freedom: 3 translations $(t_{1}, t_{2}, t_{3})$ and three rotation angles $(\alpha_{1}, \alpha_{2}, \alpha_{3})$. A 3D rigid transformation which first rotates position and then applies a translation is given by: \(T(S) = R(S) + [t_1 \; t_2 \; t_3]^T\)

where the rotation matrix $R$ is shown in equation \ref{eq:eqn1}:
\begin{equation}
\begin{split}
R = & \begin{bmatrix}
    1 & 0 & 0 \\
    0 & \cos(\alpha_1) & \sin(\alpha_1) \\
    0 & -\sin(\alpha_1) & \cos(\alpha_1)
\end{bmatrix}\\
& \begin{bmatrix}
    \cos(\alpha_2) & 0 & -\sin(\alpha_2) \\
    0 & 1 & 0 \\
    \sin(\alpha_2) & 0 & \cos(\alpha_2)
\end{bmatrix}\\
& \begin{bmatrix}
    \cos(\alpha_3) & \sin(\alpha_3) & 0 \\
    -\sin(\alpha_3) & \cos(\alpha_3) & 0 \\
    0 & 0 & 1
\end{bmatrix}
\end{split}
\label{eq:eqn1}
\end{equation}

The registration process starts with an initial alignment based on the centroids and principal axes of both images to achieve a foundational correspondence between the physical contents. Throughout the rigid registration workflow, depicted in figure~\ref{fig6}, an optimizer, a similarity metric, and an interpolator are used to refine the image alignment through an iterative process. Mattes Mutual Information \cite{neri2008image}, a metric quantifying the dependency between pixel intensities of corresponding images, is employed for image matching, utilizing a 50-bin histogram for estimation. To compute the metric, a $10 \%$ random sampling of points is used. The registration model integrates linear interpolation and a gradient descent optimizer, with a learning rate of $0.1$, a $1000$ iteration count, and a convergence threshold of $1 \times 10^{-10}$. A multi-resolution approach is adopted, altering image resolution throughout the registration to enhance robustness and accuracy. The shrink factor array $[2, 1, 1]$
 scales down the image size initially by half, maintaining the original size in subsequent levels. Gaussian smoothing, determined by the 'smoothing sigmas' parameter with values $[4, 2, 1]$, reduces noise and variations across the levels, thereby mitigating local optima during the registration process. Upon reaching convergence, both the transformation matrix and the registered image are saved.
 
\begin{figure}  
\centering
	\includegraphics[width= 3.5 in]{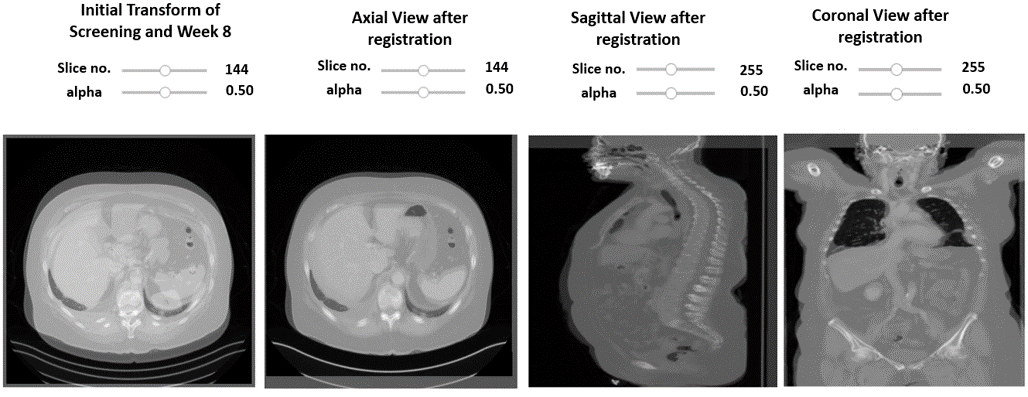}
	\caption{Overlay of the moving and fixed image after initialization (left), registered (Week 8 image after registration) and Screening (fixed) image showing the effectiveness of 3D rigid registration across three orientations.The slice number denotes the specific slice chosen within each orientation, while alpha represents the parameter controlling transparency.}\label{fig7}
\end{figure}

Figure~\ref{fig7} (left plot) displays this initial alignment, between Screening and Week 8 image series for subject 1 of Table~\ref{tab:table1}. A triaxial representation of the overlay between the registered and fixed images is subsequently presented. The congruence of the images is evident in the axial, sagittal, and coronal views, indicating a successful volumetric registration with a substantial overlap of the registered and fixed images, affirming the effectiveness of the applied registration process. Next, an inverse transformation is done on the obtained transformation matrix and this inverse transform is applied on the lesion coordinates (spatial coordinates of lesion centroids) of the moving (here Week 8) image series. Finally lesion correspondence is conducted as before between the fixed image lesion coordinates and transformed moving image lesion coordinates.

\section{Results and Discussion}\label{results}
\begin{figure}  
\centering
	\includegraphics[width= 3.5in, height =6 in]{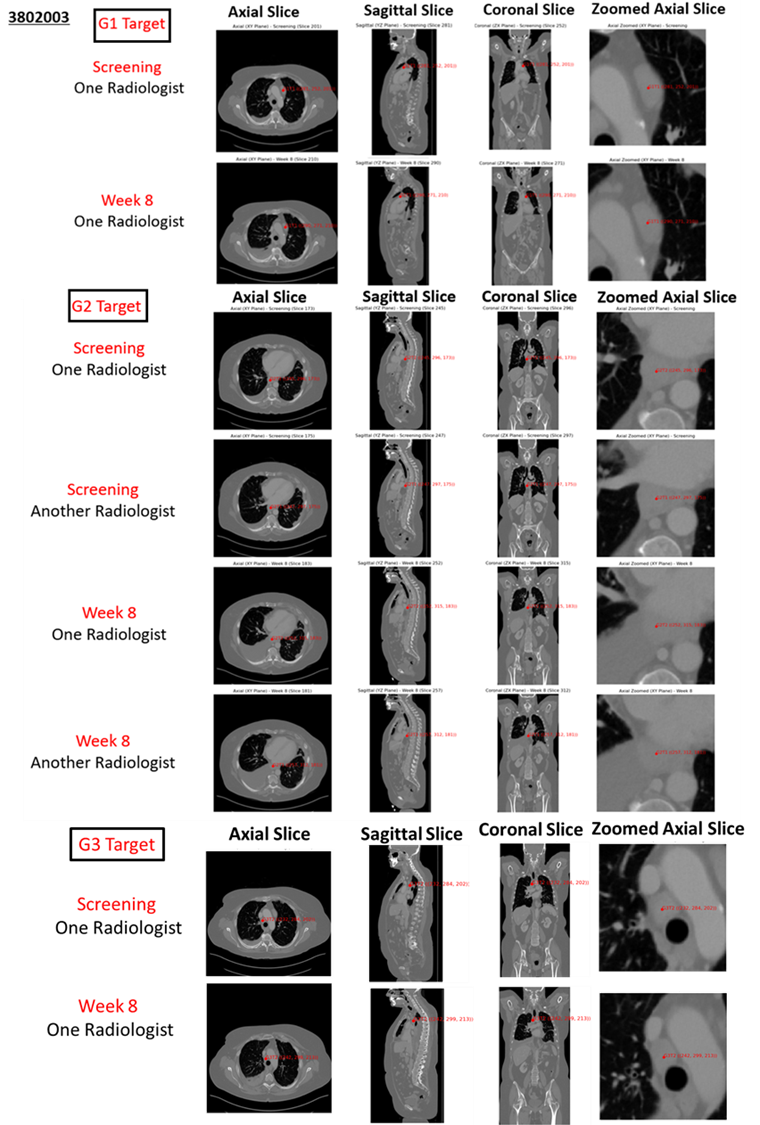}
	\caption{Visualization of the lesion correspondence algorithm performance in identifying and tracking unique target lesions (target centroids shown in red) for subject 1 of Table~\ref{tab:table1} }\label{fig8}
\end{figure}

\begin{figure} 
\centering
	\includegraphics[width= 3.5in, height = 3.5 in]{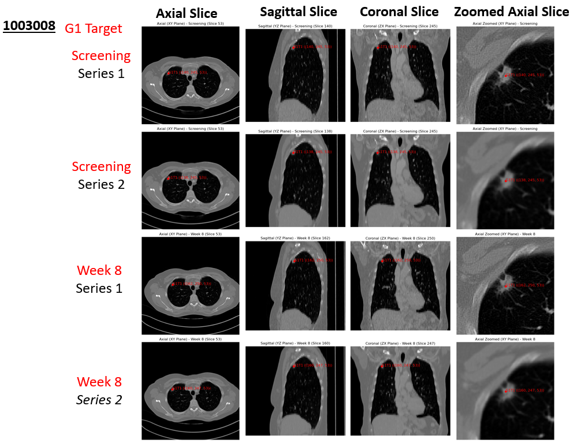}
	\caption{Visualization of the lesion correspondence algorithm performance in identifying and tracking a target lesions (target centroids shown in red) for subject 5 of Table~\ref{tab:table1}} 
 \label{fig9}
\end{figure}

We established correspondence between lesions across different timepoints, radiologists, and image scan series. For each individual patient, identified by their unique subject ID, we manually validated the correspondence and matching of lesions. This was accomplished through visual inspection of triaxial plots showing the lesion locations (see figure~\ref{fig8} and ~\ref{fig9}). Visualization ensures that the distinctiveness of each lesion is maintained, contingent on the lesion residing within the same region in the triaxial plots. We have adopted an organized approach in our triaxial plots: each column portrays a specific orientation of the lesion, while the rows represent differences in the data such as different timepoints, radiologist interpretations, or image scan series in which the lesion was detected. 

An example of our algorithm’s performance is evident in the case of the example subject. We validate our RALMAC algorithm's output focusing specifically on target lesions, highlighted in red for clarity. As depicted in Figure~\ref{fig8}, three distinct target lesions, denoted as  ${G_{1}, G_{2}}, G_{3}$ are identified. The consistency in the spatial positioning of the target lesion ‘$G_1$’ in the triaxial plots effectively underscores its uniqueness across Screening and Week 8 (see first two rows). Similarly, the lesion ‘$G_2$’, '$G_3$' demonstrates the algorithm's precision, maintaining lesion correspondence across multiple timepoints and radiologist assessments (see rows 3 to 6 for $G_2$ and rows 7-8 for $G_3$).

%\newpage % Start page 6

\vspace*{1in}

For a different subject, (figure~\ref{fig9}) our RALMAC algorithm tracked a single unique target lesion, ‘$G_1$' across different image series and timepoints. The uniformity of the lesion's location across the triaxial plots  indicates our algorithm’s robust ability to accurately establish lesion correspondence.

\begin{figure}[htbp]  
\centering
	\includegraphics[width= 3.5 in, height = 2.5 in]{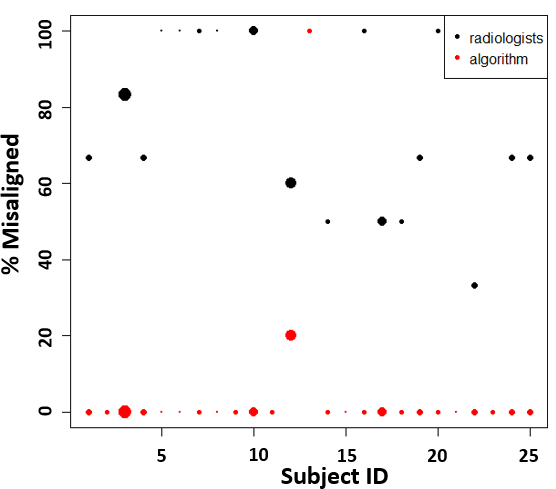}
	\caption{Plot showing the \% of misaligned target lesions reported by radiologists (in black) and the misclassification rate of the algorithm (in red) across the $25$ patients. The size of the dots are proportional to true number of target lesions.}\label{fig10}
\end{figure}

We looked at the performance of the proposed algorithm based on
screening and week 8's CT scan data from 25 randomly chosen patients
in the dataset. In Table~\ref{tab:table1}, we report the number of target lesions for each patient that were reported by two radiologists (column R1 and R2 of table~\ref{tab:table1}) as well as the number of misaligned target lesions between them. In the misaligned (M/X/N stands for Misaligned/Missing/ New) column, we not only sum the number of target lesions that had different nomenclature in the two radiologist's markings but also target lesions that were reported by one and missed by the other. Figure~\ref{fig10} show that there is a significant fraction of misalignment between the two radiologists in the data.
In columns 2, 3 and 4 of the table~\ref{tab:table1}, we respectively report the number
of unique targe lesions that are reported by our algorithm (U), the number
of target lesions that are missed by our algorithm (MI)  and the number of
target lesions that are falsely reported by our algorithm (F). 

To better conceptualize what we mean by missing and falsely reported targets, we provide examples below:
\begin{itemize}
\item \textbf{MI:} Assume that only one target $(G_{1})$ is identified by the radiologists on Screening, and two different targets were identified at Week 8. If the algorithm matches both target lesions seen on Week 8 to $G_1$, then this constitutes a missing target. 
\item \textbf{F:} Assume that a particular lesion $(G_{1})$  is identified by the radiologists on both Screening and Week 8 . If the algorithm fails to match the lesion seen on Week 8 to $G_1$ and instead calls it a new target, then this constitutes a false report.
\end{itemize}

Thus, the true
number of target lesions for each patient is $U+MI-F$.  We see that our proposed
algorithm fails for 2 out of 25 or 8\% of the patients. In both cases,
it overestimates the number of target lesions. As such our algorithm estimated a total of 65 unique targets lesions when the true number of target lesions was 62, leading to an overestimation rate of 4.84\%.

On closer introspection, we found that the overestimation of target lesions by the proposed algorithm was mainly due to a failure of aligning initially misaligned targets. The algorithm failed to recognize the misaligned targets as targets which are in close proximity and should have been merged and labeled as a single target. Overall, the algorithm worked well in labeling targets and never under-counted.

\begin{table}[htbp]
  \caption{Evaluation of the correspondence algorithm on 25 random patients in the dataset.}
  \begin{adjustbox}{width = 3.5 in, height = 2.5in}
    \begin{tabular}{lrrrrrrr}
    \toprule
    \textbf{Subject} & \multicolumn{3}{c}{\textcolor[rgb]{.439, .188, .627}{\textbf{Algorithm}}} & \multicolumn{4}{c}{\textcolor[rgb]{1, 0, 0}{\textbf{Radiologist}}} \\
    \midrule
          & \textcolor[rgb]{.439, .188, .627}{U} & \textcolor[rgb]{.439, .188, .627}{MI} & \textcolor[rgb]{.439, .188, .627}{F} & \textcolor[rgb]{1, 0, 0}{R1} & \textcolor[rgb]{1, 0, 0}{R2} & \textcolor[rgb]{1, 0, 0}{Align} & \textcolor[rgb]{1, 0, 0}{M/X/N} \\
    \midrule
    1& 3     & 0     & 0     & 2     & 2     & 1     & 2 \\
    2& 2     & 0     & 0     & 2     & 2     & 2     & 0 \\
    3& 6     & 0     & 0     & 5     & 3     & 1     & 5 \\
    4& 3     & 0     & 0     & 2     & 2     & 1     & 2 \\
    5& 1     & 0     & 0     & 1     & 0     & 0     & 1 \\
    6& 1     & 0     & 0     & 1     & 0     & 0     & 1 \\
    7& 2     & 0     & 0     & 2     & 0     & 0     & 2 \\
    8& 1     & 0     & 0     & 1     & 0     & 0     & 1 \\
    9& 2     & 0     & 0     & 2     & 2     & 2     & 0 \\
    10& 4     & 0     & 0     & 2     & 2     & 0     & 4 \\
    11& 2     & 0     & 0     & 2     & 2     & 2     & 0 \\
    12& 6     & 0     & 1     & 5     & 2     & 2     & 3 \\
    13& 4     & 0     & 2     & 2     & 0     & 0     & 2 \\
    14& 2     & 0     & 0     & 2     & 1     & 1     & 1 \\
    15& 1     & 0     & 0     & 1     & 1     & 1     & 0 \\
    16& 2     & 0     & 0     & 2     & 0     & 0     & 2 \\
    17& 4     & 0     & 0     & 4     & 2     & 2     & 2 \\
    18& 2     & 0     & 0     & 2     & 1     & 1     & 1 \\
    19& 3     & 0     & 0     & 2     & 2     & 1     & 2 \\
    20& 2     & 0     & 0     & 1     & 1     & 0     & 2 \\
    21& 1     & 0     & 0     & 1     & 0     & 0     & 1 \\
    22& 3     & 0     & 0     & 3     & 2     & 2     & 1 \\
    23& 2     & 0     & 0     & 2     & 2     & 2     & 0 \\
    24& 3     & 0     & 0     & 1     & 3     & 1     & 2 \\
    25& 3     & 0     & 0     & 3     & 2     & 1     & 2 \\
    \bottomrule
    \end{tabular}%
    \end{adjustbox}%
  \label{tab:table1}%
\end{table}%

\section{Conclusion and Future Work}\label{conclusion}
This paper introduces a novel registration-based automated lesion correspondence and matching (RALMAC) algorithm which was shown to works well in tracking target lesions and non-target lesions in diverse CT imaging datasets based on our preliminary assessment of 25 patients. We plan to make our implementation publicly available as a CDRH regulatory science tool \cite{cdrh2023rst}. As part of our future work, we aim to assess the efficacy of the algorithm in lesion tracking across all follow-up timepoints (till the end of follow up or an event). The establishment of lesion correspondence across the timepoints will be instrumental for the next phase of our overall study that aims to predict patient response to therapy based on longitudinal imaging. 

\section*{Acknowledgment}
This work is supported by the FDA Office of Women’s Health. Subrata Mukherjee’s appointment was supported by the Research Participation Program at the U.S. Food and Drug Administration administered by the Oak Ridge Institute for Science and Education through an interagency agreement between the U.S. Department of Energy and the U.S. Food and Drug Administration.

\clearpage

\bibliographystyle{ieeetr}
\bibliography{mybib.bib}

\begin{thebibliography}{10}

\bibitem{coroller2023methodology}
T.~Coroller, B.~Sahiner, A.~Amatya, A.~Gossmann, K.~Karagiannis, C.~Moloney, R.~K. Samala, L.~Santana-Quintero, N.~Solovieff, C.~Wang, {\em et~al.}, ``Methodology for good machine learning with multi-omics data,'' {\em Clinical Pharmacology \& Therapeutics}, 2023.

\bibitem{coroller2023multi}
T.~P. Coroller, B.~Sahiner, A.~Amatya, A.~Gossman, K.~Karagiannis, R.~K. Samala, L.~Santana-Quintero, N.~Solovieff, C.~Wang, L.~Amiri-Kordestani, {\em et~al.}, ``Multi-omics investigation on the prognostic and predictive factors in metastatic breast cancer using data from phase iii ribociclib clinical trials: A statistical and machine learning analysis plan,'' {\em medRxiv}, pp.~2023--08, 2023.

\bibitem{tripathy2018ribociclib}
D.~Tripathy, S.-A. Im, M.~Colleoni, F.~Franke, A.~Bardia, N.~Harbeck, S.~A. Hurvitz, L.~Chow, J.~Sohn, K.~S. Lee, {\em et~al.}, ``Ribociclib plus endocrine therapy for premenopausal women with hormone-receptor-positive, advanced breast cancer (monaleesa-7): a randomised phase 3 trial,'' {\em The Lancet Oncology}, vol.~19, no.~7, pp.~904--915, 2018.

\bibitem{slamon2018phase}
D.~J. Slamon, P.~Neven, S.~Chia, P.~A. Fasching, M.~De~Laurentiis, S.-A. Im, K.~Petrakova, G.~V. Bianchi, F.~J. Esteva, M.~Mart{\'\i}n, {\em et~al.}, ``Phase iii randomized study of ribociclib and fulvestrant in hormone receptor--positive, human epidermal growth factor receptor 2--negative advanced breast cancer: Monaleesa-3,'' {\em Journal of Clinical Oncology}, vol.~36, no.~24, pp.~2465--2472, 2018.

\bibitem{hortobagyi2016ribociclib}
G.~N. Hortobagyi, S.~M. Stemmer, H.~A. Burris, Y.-S. Yap, G.~S. Sonke, S.~Paluch-Shimon, M.~Campone, K.~L. Blackwell, F.~Andr{\'e}, E.~P. Winer, {\em et~al.}, ``Ribociclib as first-line therapy for hr-positive, advanced breast cancer,'' {\em New England journal of medicine}, vol.~375, no.~18, pp.~1738--1748, 2016.

\bibitem{chaffer2011perspective}
C.~L. Chaffer and R.~A. Weinberg, ``A perspective on cancer cell metastasis,'' {\em science}, vol.~331, no.~6024, pp.~1559--1564, 2011.

\bibitem{humbert2020dissociated}
O.~Humbert and D.~Chardin, ``Dissociated response in metastatic cancer: an atypical pattern brought into the spotlight with immunotherapy,'' {\em Frontiers in Oncology}, vol.~10, p.~566297, 2020.

\bibitem{wang2012study}
C.~Wang and Y.~Shen, ``Study on the distribution features of bone metastases in prostate cancer,'' {\em Nuclear medicine communications}, vol.~33, no.~4, pp.~379--383, 2012.

\bibitem{huff2023performance}
D.~T. Huff, V.~Santoro-Fernandes, S.~Chen, M.~Chen, C.~Kashuk, A.~J. Weisman, R.~Jeraj, and T.~G. Perk, ``Performance of an automated registration-based method for longitudinal lesion matching and comparison to inter-reader variability,'' {\em Physics in Medicine \& Biology}, vol.~68, no.~17, p.~175031, 2023.

\bibitem{an2021robust}
Z.~An, H.~Ma, L.~Liu, Y.~Wang, H.~Lu, C.~Zhou, R.~Xiong, and J.~Hu, ``Robust orthogonal-view 2-d/3-d rigid registration for minimally invasive surgery,'' {\em Micromachines}, vol.~12, no.~7, p.~844, 2021.

\bibitem{xu2011automated}
J.~Xu, H.~Greenspan, S.~Napel, and D.~L. Rubin, ``Automated temporal tracking and segmentation of lymphoma on serial ct examinations,'' {\em Medical physics}, vol.~38, no.~11, pp.~5879--5886, 2011.

\bibitem{santoro2021development}
V.~Santoro-Fernandes, D.~Huff, M.~L. Scarpelli, T.~G. Perk, M.~R. Albertini, S.~Perlman, S.~S. Yip, and R.~Jeraj, ``Development and validation of a longitudinal soft-tissue metastatic lesion matching algorithm,'' {\em Physics in Medicine \& Biology}, vol.~66, no.~15, p.~155017, 2021.

\bibitem{muenzel2012intra}
D.~Muenzel, H.-P. Engels, M.~Bruegel, V.~Kehl, E.~Rummeny, and S.~Metz, ``Intra-and inter-observer variability in measurement of target lesions: implication on response evaluation according to recist 1.1,'' {\em Radiology and oncology}, vol.~46, no.~1, pp.~8--18, 2012.

\bibitem{munkres1957algorithms}
J.~Munkres, ``Algorithms for the assignment and transportation problems,'' {\em Journal of the society for industrial and applied mathematics}, vol.~5, no.~1, pp.~32--38, 1957.

\bibitem{mills2007dynamic}
G.~A. Mills-Tettey, A.~Stentz, and M.~B. Dias, ``The dynamic hungarian algorithm for the assignment problem with changing costs,'' {\em Robotics Institute, Pittsburgh, PA, Tech. Rep. CMU-RI-TR-07-27}, 2007.

\bibitem{marstal2016simpleelastix}
K.~Marstal, F.~Berendsen, M.~Staring, and S.~Klein, ``Simpleelastix: A user-friendly, multi-lingual library for medical image registration,'' in {\em Proceedings of the IEEE conference on computer vision and pattern recognition workshops}, pp.~134--142, 2016.

\bibitem{larrey2006optimal}
J.~Larrey-Ruiz and J.~Morales-S{\'a}nchez, ``Optimal parameters selection for non-parametric image registration methods,'' in {\em International Conference on Advanced Concepts for Intelligent Vision Systems}, pp.~564--575, Springer, 2006.

\bibitem{neri2008image}
E.~Neri, D.~Caramella, C.~Bartolozzi, {\em et~al.}, ``Image processing in radiology,'' {\em Medical Radiology. Diagnostic Imaging. Springer, Berlin}, 2008.

\bibitem{cdrh2023rst}
CDRH, ``Catalog of regulatory science tools to help assess new medical devices,'' 2023.
\newblock \url{https://www.fda.gov/medical-devices/science-and-research-medical-devices/catalog-regulatory-science-tools-help-assess-new-medical-devices} Accessed: Feb. 15th, 2024.

\end{thebibliography}

\end{document}